\documentclass[10pt,twocolumn,prb,aps,amssymb,amsmath,superscriptaddress]{revtex4}
\include{gyrokinetics-macros}
\usepackage{color,graphicx,ulem,bm}

\newcommand{\mbf}[1]{\mathbf{#1}}
\newcommand{\lb}{\left<}
\newcommand{\rb}{\right>}
\newcommand{\parl}{\parallel}
\newcommand{\pd}[2]{\frac{\partial #1}{\partial #2}}
\newcommand{\unit}[1]{\mathbf{\hat{#1}}}

\renewcommand{\eqref}[1]{Eq.\ (\ref{#1})}

\newcommand{\secref}[1]{Sec.\ \ref{#1}}

\newcommand{\tld}[1]{\tilde{#1}}
\newcommand{\avg}[1]{\lb#1\rb}

\newcommand{\vvol}{d^{3}\mbf{v}}

\newcommand{\vpa}{v_{\parallel}}
\newcommand{\vpe}{v_{\perp}}

\newcommand{\beq}{\begin{equation}}
\newcommand{\eeq}{\end{equation}}
\newcommand{\grad}{\nabla}

\begin{document}

\title{Direct multiscale coupling of a transport code to gyrokinetic turbulence codes}

\author{M.\ Barnes}
\email{michael.barnes@physics.ox.ac.uk}
\affiliation{
Rudolf Peierls Centre for Theoretical Physics, University of Oxford, 1 Keble Road, Oxford OX1 3NP, UK
}
\affiliation{
Euratom/CCFE Fusion Association, Culham Science Centre, Abingdon OX14 3DB, UK
}
\author{I.\ G.\ Abel}
\affiliation{
Rudolf Peierls Centre for Theoretical Physics, University of Oxford, 1 Keble Road, Oxford OX1 3NP, UK
}
\affiliation{
Euratom/CCFE Fusion Association, Culham Science Centre, Abingdon OX14 3DB, UK
}
\author{W.\ Dorland}
\affiliation{
Department of Physics, University of Maryland, College Park, Maryland 20742-3511
}
\author{T.\ G\"orler}
\affiliation{
Max-Planck-Institut f\"ur Plasmaphysik, Boltzmannstrasse 2, D-85748 Garching, Germany
}
\author{G.\ W.\ Hammett}
\affiliation{
Princeton Plasma Physics Laboratory, Princeton University, P.O. Box 451, Princeton, New Jersey 08543
}
\author{F.\ Jenko}
\affiliation{
Max-Planck-Institut f\"ur Plasmaphysik, Boltzmannstrasse 2, D-85748 Garching, Germany
}

\begin{abstract}

Direct coupling between a transport solver and local, nonlinear gyrokinetic calculations using the multiscale gyrokinetic code \texttt{TRINITY} [M. Barnes, Ph.D. thesis, arxiv:0901.2868] is described.  The coupling of the microscopic and macroscopic physics is done within the framework of multiscale gyrokinetic theory, of which we present the assumptions and key results.  An assumption of scale separation in space and time allows for the simulation of turbulence in small regions of the space-time grid, which are embedded in a coarse grid on which the transport equations are implicitly evolved.  This leads to a reduction in computational expense of several orders of magnitude, making first-principles simulations of the full fusion device volume over the confinement time feasible on current computing resources.  Numerical results from \texttt{TRINITY} simulations are presented and compared with experimental data from JET and ASDEX Upgrade plasmas.

\end{abstract}



\maketitle

\section{Introduction}
\label{sec:intro}

A fundamental challenge of fusion science is to maximize fusion power, which is determined primarily by macroscopic profiles of density and temperature.  These profiles, which vary spatially on the system scale and evolve on the energy confinement time scale, drive turbulence at micro-scales in space and time.  In the absence of MHD instability, this micro-turbulence is the dominant source of heat flux observed in standard tokamaks, which sets rigid constraints on the macroscopic profiles.~\cite{kotschPoP95}  Consequently, it is of critical importance to understand the self-consistent interaction between the macroscopic profiles and the micro-turbulence.

This is a challenging task due to both the wide range of scales involved and the high dimensionality of the system.  The electron turbulence space scale, which is comparable to the electron Larmor radius, will be on the order of $0.1$ millimeters for a device such as ITER,~\cite{aymarNF01} which has a minor radius of about two meters.  Similarly, the electron turbulence time scale is approximately one microsecond, much smaller than the expected energy confinement time of several seconds.  Additionally, the instabilities driving the turbulence are kinetic in nature, requiring treatment of the velocity space in addition to the configuration space.  Including all of these dynamics directly in a single simulation is not feasible on current computing resources.

As long as all relevant dynamics occur on time scales long compared to particle gyro-motion, it is possible to average over the gyro-orbits and eliminate the gyro-angle as a phase space variable.  Furthermore, for magnetic confinement devices with sufficiently small $\rho_*$ (ratio of ion Larmor radius to plasma minor radius), there is expected to be a separation between micro- and macro-scales in space and time.~\cite{iterNF99,mcdonaldIAEA04}  The $\delta f$-gyrokinetic model~\cite{antonPoF80,friemanPoF82,howesApJ06} exploits these scale separations, simplifying the system considerably.  Given a set of fixed macroscopic profiles, it allows for the calculation of turbulent fluxes in a $5$-D phase space.  Such calculations have been performed numerically in gyrokinetic codes for more than two decades, providing much insight into the nature of kinetic instabilities and micro-turbulence.   First-principles $\delta f$-gyrokinetic simulations have steadily advanced in their sophistication and physical fidelity and have become routine in recent years.  

However, these codes only model the effect of macroscopic profiles on micro-turbulence: They do not provide quantitative information on how this turbulence subsequently affects the evolution of the macroscopic profiles.  Accurately calculating fluxes from experimental profiles is also problematic because of the acute sensitivity of the fluxes to small changes in the input profile gradients (which would arise due to uncertainty in experimental measurements).~\cite{candyPRL03}  In an attempt to address these issues, full-$f$ gyrokinetic codes have been developed, which do not explicitly assume the scale separations listed above.  Subsequently, they are able to take into account the two-way interaction between turbulence and equilibrium thermodynamic profiles.  However, well-resolved full-$f$ simulations would be extremely expensive numerically because of the wide range of scales described above and because of the necessity of calculating the distribution function and fields to very high order.  It has also recently been argued that this approach as currently formulated leads to an unphysical source of toroidal angular momentum.~\cite{parraPPCF08}

An alternative approach commonly used to study the interaction of the micro- and macro-physics is to solve fluid transport equations with a reduced model for the turbulent fluxes.  Typically, the transport is modeled as a diffusive process, with turbulent diffusivities coming from a wide range of models, including empirical fits to experiment or simulation and theory-based estimates.~\cite{nordmanNF90,kotschPoP95,kinseyPoP96,waltzPop97,staeblerPoP05,bourdellePoP07}  These reduced models have provided a basic qualitative understanding of the multiscale interaction and are capable in some cases of giving good quantitative agreement with first-principles, nonlinear gyrokinetic calculations.~\cite{kinseyPoP08,casatiNF09}  However, such models do not permit detailed validation studies because they do not produce fluctuation spectra or other data that can be experimentally challenged.  Without careful validation, even first-principles reduced models (with no adjustable parameters) cannot be fully trusted for predictions of performance in new operational regimes, and reduced models with fit parameters adjusted for current experiments have even less credibility.  There are known cases for which even first-principles reduced models currently available are inadequate.~\cite{staeblerPPCF04,candyIAEA08}  


Consequently, it is desirable to couple not only reduced flux models, but also direct numerical simulations of the turbulence, to transport solvers.  This has been done in Ref.~\onlinecite{shestakovJCP03} for a Hasegawa-Wakatani 2D fluid model of the turbulence using an implicit relaxation technique and allowing for nonlocality.  Here, we describe coupling to local, nonlinear, 5D gyrokinetic turbulence calculations using a Newton method (similar to that described in Ref.~\onlinecite{jardinJCP08}), which accelerates convergence by more than an order of magnitude for typical parameters.  This coupling is achieved using the multiscale gyrokinetic code, \texttt{TRINITY},~\cite{barnes} which can use nonlinear fluxes from the continuum gyrokinetic codes \texttt{GS2}~\cite{kotschCPC95} or \texttt{GENE}.~\cite{jenkoPoP00,gene}  Our approach is similar to that employed in \texttt{TGYRO},~\cite{candyPoP09} with the key distinction that \texttt{TRINITY} evolves the macroscopic profiles in time, whereas \texttt{TGYRO} assumes a steady-state and solves the volume-integrated transport equations for profile gradients.  

It should be noted that some meso-scale phenomena in space and time are not formally considered in the standard multiscale gyrokinetic model. Our simulations ignore low-order magnetic islands and so are directly applicable only during MHD-quiescent periods of plasmas. Rapid cold/heat pulse propagation (such as following a sawtooth or ELM crash) is possible in the \texttt{TRINITY} code because of the presence of stiff critical gradients in ITG and TEM turbulence, though the transport time step would of course have to be reduced in \texttt{TRINITY} during such a transient event to be able to follow its propagation.  It should be noted that flux-tube simulations include the contribution to the heat flux of avalanches on all scales up to the radial size of the flux tubes. If even longer wavelength avalanches were important, then the heat flux would increase as the flux-tube simulation domain was made larger, so convergence studies can be used to test this.  PIC and continuum flux-tube simulations generally find that the flux converges with sufficiently large simulation size, of order the sizes we are using here. Previous gyrokinetic studies have found that some modest non-local turbulence spreading may occur over distances of a few radial eddy sizes,~\cite{hahmPPCF04} but in the core region of the large tokamaks we are studying here (and even more so at reactor scales) this should usually be a small effect.  It should be acknowledged that the separation of scales assumed for the core plasma in this paper may break down in the edge region of the plasma because gradient scale lengths and eddy sizes may not be very different near the edge, so non-local effects may be important there.

This paper is organized as follows: in the next section we state the fundamental assumptions of the multiscale model and present the closed system of equations that results from gyrokinetic expansion of the Maxwell-Boltzmann system.  In \secref{sec:numerics}, we describe the numerical scheme used in \texttt{TRINITY} to simulate the multiscale gyrokinetic system of equations.  We also give estimates for the space and time domain savings provided by the multiscale scheme.  \secref{sec:results} contains results from \texttt{TRINITY} simulations of L-mode and H-mode discharges from JET and ASDEX Upgrade.  We show that the numerical data from \texttt{TRINITY} is in good quantitative agreement with reconstructions of experimental data.  Finally, we conclude in \secref{sec:conclusion} with a summary and a discussion of possible future directions for research.

\section{Theoretical framework}
\label{sec:theory}

In this section, we state the fundamental assumptions present in our multiscale model and present the resulting closed system of equations that must be solved.  These equations, which are a rigorous asymptotic limit of the full Maxwell-Boltzmann system, have been derived in detail in Refs.~\onlinecite{sugamaPoP97} and~\onlinecite{abelPoP09}.  We include a brief overview of the key results here for completeness.

As the starting point for our analysis, we begin with the coupled system consisting of Maxwell's equations and the driven Fokker-Planck equation:
\beq
\frac{df_s}{dt}=C[f_s]+S_s[f_s],
\label{eqn:fp}
\eeq
where $f_s=f_s(\mbf{x},\mbf{v},t)$ represents the distribution of particles of species $s$ in position ($\mbf{x}$) and velocity ($\mbf{v}$) space, $C$ represents the effect of two-particle Coulomb interactions, and $S_s$ is a source term accounting for the external injection of particles, momentum, and energy.  This system of equations describes all of the important dynamics in fusion plasmas and is consequently intractable, both analytically and numerically.  Since we are interested in studying the interaction of the plasma micro-turbulence with the macroscopic profiles, we simplify the system by adopting a variant of the standard $\delta f$-gyrokinetic ordering.~\cite{friemanPoF82}  This ordering imposes constraints on the relative amplitudes and space-time scales of the macro- and micro-physics.

We first decompose the distribution function into macroscopic and microscopic quantities by ensemble averaging: $f=\avg{f}+\delta f$, with the angled brackets denoting an ensemble average.  Defining the smallness parameter to be $\epsilon\equiv\rho/L$, where $\rho$ is the Larmor radius and $L$ is a macroscopic scale length, we order each of the terms in the Maxwell-Boltzmann equations.  The assumptions employed in ordering the terms are as follows: (1) The fluctuations are assumed to be low amplitude compared to macroscopic quantities, such that $\delta f \sim \epsilon \avg{f}$.  This is in good agreement with core measurements from a number of modern fusion experiments, which find density and temperature fluctuations $\lesssim 1\%$ of the macroscopic densities and temperatures;~\cite{fonckPRL93,evensenNF98,whitePoP08}  (2) The micro-turbulence is assumed to be spatially anisotropic with macro-scale variations along and micro-scale (i.e. Larmor radius) variations across the equilibrium magnetic field.  Experimental measurements of turbulence parallel and perpendicular correlation lengths support this hypothesis;~\cite{fonckPRL93,mckeeNF01,conwayPPCF08}  (3) All frequencies of interest are assumed to be well below the ion cyclotron frequency, and the evolution of the macroscopic profiles is taken to be much slower than the turbulent fluctuations ($\partial \avg{f}/\partial t \sim \epsilon \partial \delta f/\partial t$).  Again, this transport ordering is in agreement with experimental evidence;~\cite{mckeeNF01} (4) We order $\partial \delta f/\partial t\sim C[\delta f]$, with the collision frequency, $\nu$, defined such that $\nu \gtrsim \epsilon \partial \ln \delta f/\partial t$.  Consequently, $\delta f$ is allowed characteristic scales in velocity space, $\delta v$, of size $\sqrt{\epsilon}v_{th}\lesssim \delta v \lesssim v_{th}$.  This collision frequency ordering is satisfied even in very collisionless plasmas such as anticipated in ITER; (5) Macroscopic flows are assumed to be comparable to the ion thermal speed, with microscopic flows (i.e. $\mbf{E}\times \mbf{B}$ velocity) taken to be much smaller.  For simplicity, we consider in this paper the case in which the Mach number of the macroscopic flow is taken to be small as a subsidiary ordering; (6) The external particle, momentum, and energy sources are assumed to affect the system evolution on the confinement time scale, consistent with experiment.

Expanding the distribution function and fields in $\epsilon$ and applying the above ordering assumptions to the Maxwell-Boltzmann system results in a hierarchy of equations that is ultimately closed by ensemble and flux surface averaging and taking moments of the evolution equation for the lowest order (macroscopic) distribution function, $f_0=\avg{f_0}$.  One finds that $f_0$ is a gyroangle-independent, shifted Maxwellian whose evolution is governed by the following transport equations:
{\setlength\arraycolsep{0.1em}
\begin{eqnarray}
\label{eqn:n0}
\pd{n_s}{t} &+& \frac{1}{V'}\pd{}{\psi}\left(V'\overline{\avg{\Gamma_s}}\right) = \overline{\avg{S_n}}\\
\label{eqn:L0}
\pd{\overline{L}}{t} &+& \sum_s\frac{1}{V'}\pd{}{\psi}\left(V'\overline{\avg{\pi_s}}\right) \nonumber \\
&=& \frac{1}{4\pi}\overline{\grad\cdot\avg{\delta\mbf{B}\delta\mbf{B}\cdot\grad\phi R^2}}+ \ \sum_s \overline{\avg{S_{L_s}}}\\
\label{eqn:p0}
\frac{3}{2}\pd{p_s}{t} &+& \frac{1}{V'}\pd{}{\psi}\left(V'\overline{\avg{Q_s}}\right) \nonumber \\
&=& - \overline{\avg{H_s}} + \frac{3}{2} n_s \sum_u \nu_{su}^{\varepsilon} \left(T_u - T_s\right) + \overline{\avg{S_p}},
\end{eqnarray}
}
where $R$ is the major radius, $\phi$ is the physical toroidal angle, $\psi$ is the flux label, the overline denotes a flux surface average, and $V'=dV/d\psi$, with $V$ being the volume enclosed by the flux surface.
The evolved quantities $n_s$, $p_s$, and $L$ are the ensemble-averaged density, pressure, and species-summed toroidal angular momentum, respectively.  In the low Mach limit we are considering, the density and pressure are constant on flux surfaces.  Terms denoted by $S$ represent external sources, with subscripts indicating the relevant injected quantity.  The collisional energy exchange frequency, $\nu_{su}^{\varepsilon}$ is given in Ref.~\onlinecite{nrl}.  The terms $\Gamma$, $\pi$, $Q$, and $H$ are fluxes and heating generally consisting of classical, neoclassical, and turbulent contributions.  They are given by
{\setlength\arraycolsep{0.2em}
\begin{eqnarray}
\label{eqn:pflx}
\Gamma &\equiv& \grad\psi \cdot \int\vvol \left(\mbf{v}_{\chi} \delta f_1 + \mbf{v}_B \avg{f_1} + \bm{\rho}C\left[\bm{\rho}\cdot\grad f_0\right]\right)\\
\label{eqn:uflx}
\pi&\equiv& \grad\psi \cdot \int \vvol \ \left(mR^2\mbf{v}\cdot\grad\phi\right) \mbf{v}_{\chi} \delta f_1
\end{eqnarray}
\beq
\label{eqn:qflx}
Q\equiv\grad\psi \cdot \int \vvol \left(\mbf{v}_{\chi} \delta f_1 + \mbf{v}_B\avg{f_1} + \bm{\rho}C[\bm{\rho}\cdot\grad f_0]\right)\frac{mv^2}{2}
\eeq
\beq
\label{eqn:heat}
H\equiv \int \vvol \ e\left(\frac{D\chi}{Dt} + \pi \pd{\omega}{\psi} + \frac{\mbf{v}}{c}\cdot \pd{\mbf{A}_0}{t}\left(\avg{f_1}+\bm{\rho}\cdot\grad f_0\right)\right),
\eeq
}
where $\chi=\delta \Phi-\mbf{v}\cdot\delta\mbf{A}/c$ is the generalized electromagnetic potential fluctuation, $\mbf{v}_{\chi}=c/B\unit{b}\times\grad\chi$ is the particle drift due to the fluctuating fields, $\mbf{v}_B=(\unit{b}/\Omega)\times\left(\vpe^2\grad \ln B / 2 + \vpa^2\unit{b}\cdot\grad\unit{b} \right)$ contains the magnetic drifts, $e$ is the species charge, $\bm{\rho}$ is the gyroradius vector, and $D/Dt = \partial/\partial t + \mbf{u}\cdot\grad$, with $\mbf{u}=R\omega(\psi)\unit{\phi}$ the equilibrium flow velocity.

The components of the first order distribution function, $f_1$, and the fluctuating potentials are obtained by solving the neoclassical and gyrokinetic equations, coupled to the low-frequency Maxwell's equations.  These equations are all self-consistently obtained as part of the multiscale gyrokinetic expansion.  The neoclassical equation governing $\avg{f_1}$ is
\beq
C[\avg{f_1}] - \vpa\unit{b}\cdot\grad \avg{f_1} = \mbf{v}_B \cdot \nabla f_0 + \frac{ef_0}{cT}\vpa\unit{b}\cdot\pd{\mbf{A}_0}{t}. 
\label{eqn:neo}
\eeq
The gyrokinetic equation determining the evolution of $\delta f_1$ is
\beq
\begin{split}
&\frac{Dh}{Dt_{\mbf{R}}} + \left(\vpa\unit{b} + \lb\mbf{v}_{\chi}\rb_{\mbf{R}} + \mbf{v}_B\right)\cdot\grad h -\lb C[h] \rb_{\mbf{R}}  \\
&= \frac{f_0}{T}\left(\frac{D\lb\chi\rb_{\mbf{R}}}{Dt_{\mbf{R}}} +\lb\mbf{v}_{\chi}\rb_{\mbf{R}}\cdot\grad\psi\left(m\vpa\frac{I}{B}\pd{\omega}{\psi} - \pd{\ln f_0}{\psi} \right)\right),
\label{eqn:gk}
\end{split}
\eeq
where $\delta f_1 = h-(q\delta\Phi/T) f_0$, $I(\psi)=q(\psi)/(\overline{1/R^2})$ is the toroidal flux function, $q(\psi)$ is the safety factor, $\lb . \rb_{\mbf{R}}$ denotes a gyroaverage at fixed guiding center position, $\mbf{R}$, and the subscript on the $D/Dt$ operator indicates that it is to be evaluated at $\mbf{R}$.  The low-frequency Maxwell's equations are given by
\begin{eqnarray}
\label{eqn:poisson}
\grad_{\perp}^2 \delta\phi &=& -4\pi\sum_s e_s \int\vvol  \ h_s\\
\label{eqn:paramp}
\grad_{\perp}^2 \delta A_{\parl} &=& -\frac{4\pi}{c}\sum_s e_s \int\vvol \ \vpa h_s\\
\label{eqn:perpamp}
\grad_{\perp} \delta B_{\parl} &=& \frac{4\pi}{c}\sum_s e_s \int\vvol \left(\unit{b}\times\mbf{v}_{\perp}\right) h_s.
\end{eqnarray}

\begin{figure*} 
\begin{center}
\includegraphics[height=7.0cm,angle=0]{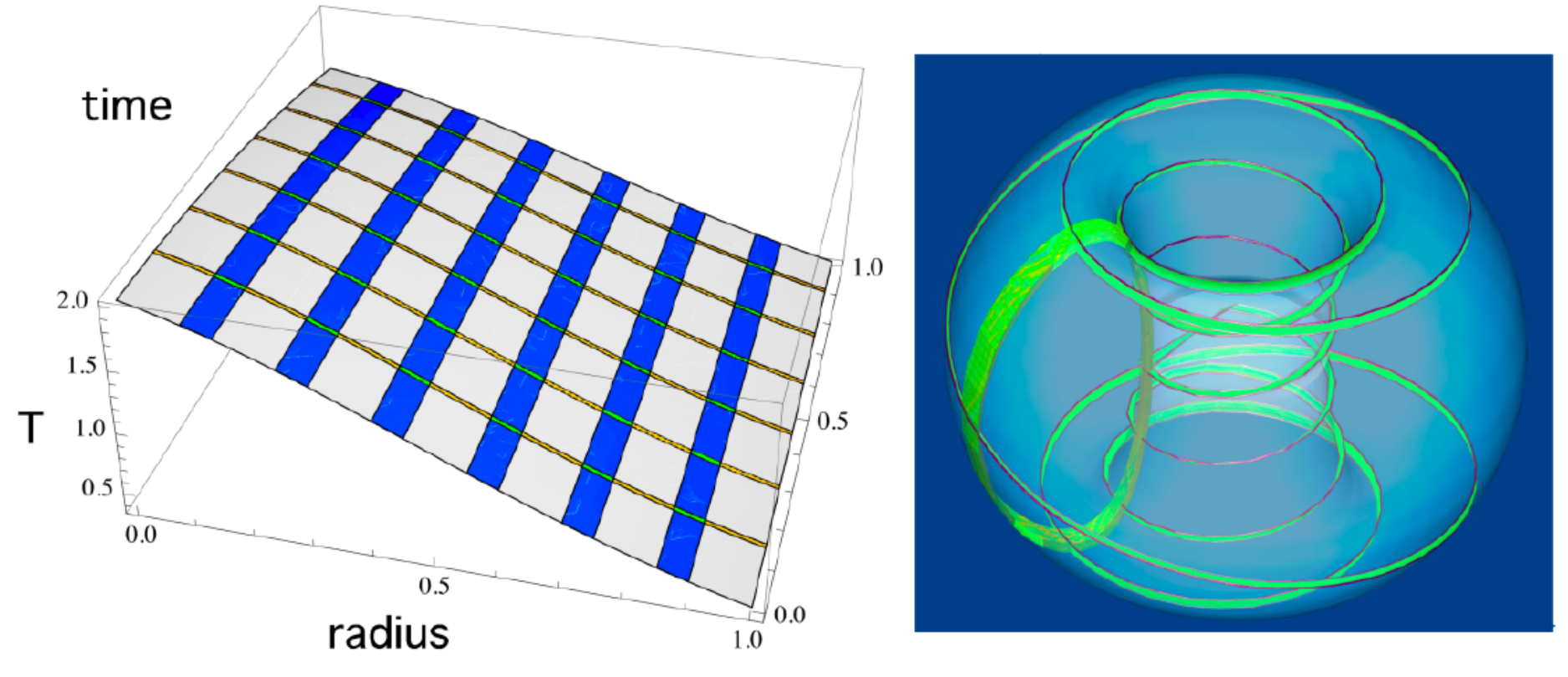}
\end{center}
\caption{(Left:) Cartoon of multiscale space-time grid used in \texttt{TRINITY}.  Blue (vertical) and yellow (horizontal) regions represent the radial and time domains, respectively, over which a fine mesh is used to calculate turbulent fluxes.  The overlapping regions, colored in green, denote the reduced space-time domain used in \texttt{TRINITY}.  These green patches, each of which corresponds to a nonlinear gyrokinetic flux tube simulation (right), are grid points in the coarse space-time mesh used to solve the transport equations.}
\label{fig:msgrid}
\end{figure*}

The final elements needed to close this system are equations for the evolution of the magnetic geometry.  In particular, the magnetic field can be specified in an axisymmetric system if given the poloidal flux, $\psi$, and the toroidal flux function, $I(\psi)$.  The toroidal flux function is evolved by taking the toroidal component of Faraday's Law and flux surface averaging:
\beq
\pd{}{t}\overline{\left(\frac{I(\psi)}{R^2}\right)} = -\frac{1}{V'}\pd{}{\psi}\left(V' \overline{\frac{\mbf{B}}{c}\cdot\pd{\mbf{A}_0}{t}}\right),
\label{eqn:qpsi}
\eeq
with the $\overline{\mbf{B}\cdot\partial \mbf{A}_0/\partial t}$ term obtainable from the neoclassical equation (Eq.~\ref{eqn:neo}).  Eqs.~\ref{eqn:neo} and~\ref{eqn:qpsi} are then coupled to the Grad-Shafranov equation, which uses the updated macroscopic pressure from Eq.~\ref{eqn:p0} to obtain $\psi$ and close the system:
\beq
R^2 \grad\cdot\left(\frac{\grad\psi}{R^2}\right) = -I\pd{I}{\psi} - 4\pi R^2 \sum_s \pd{p_s}{\psi}.
\label{eqn:gradshaf}
\eeq

\section{Numerical method}
\label{sec:numerics}

We describe in this section the numerical model we have developed for solving the system described by Eqs.~\ref{eqn:n0}-\ref{eqn:gradshaf}.  This model is implemented in the multiscale gyrokinetic transport solver \texttt{TRINITY}.~\cite{barnes}  Currently, there are a few additional assumptions employed in \texttt{TRINITY} to simplify the multiscale system presented in the previous section.  In what follows, we provide a numerical prescription for solving the full system, pointing out the places where the \texttt{TRINITY} model has been simplified.

A simple sketch of our multiscale numerical model is given in Fig.~\ref{fig:msgrid}.  A direct numerical simulation would require us to have a fine space-time mesh over the full device volume and over at least a confinement time.  We exploit scale separation present in the system to drastically reduce the domain over which a fine mesh is required.  Our assumption of time scale separation between the turbulence and the equilibrium allows us to fix equilibrium quantities while we evolve the turbulence to saturation.  Additionally, it allows us to use steady-state, time-averaged fluxes in our transport equations.  Consequently, we need only resolve turbulence time scales for short periods of time, between which we can take large time steps characteristic of the confinement time.  As the separation of scales gets wider, the simulation domain savings from this approach grows: The cost of simulating small $\rho_*$ devices is no greater than that for moderate $\rho_*$ devices.  The time domain savings for a device like ITER is a factor of hundreds.

Similarly, spatial scale separation allows us to assume that macroscopic quantities (and their associated gradient scale lengths) are constant across the radial domain in which we simulate turbulent dynamics.  As long as the turbulence simulation domain is wide enough in each dimension, the turbulence at the ends of the domain is uncorrelated.  Statistically periodic boundary conditions then apply.  The result of this local approximation is a flux tube simulation domain for the turbulence (Fig.~\ref{fig:msgrid}), which can be used to periodically map out a flux surface.  Comparisons between local and global gyrokinetic simulations have shown that the local approximation is valid for small $\rho_*$,~\cite{candyPop04} as it must be for the gyro-Bohm scaling suggested by high confinement experiments~\cite{mckeeNF01} to hold.  Once again, the spatial domain savings increases with the scale separation.  On small $\rho_*$ devices, the simulation volume is reduced by approximately a factor of one hundred.

\subsection{Discretization of the transport equations}
\label{sec:discrete}

The transport equations (Eqs.~\ref{eqn:n0}-\ref{eqn:p0}) are stiff, nonlinear partial differential equations.  In order to take the large time steps required by our multiscale scheme, we must treat them implicitly.  We allow for a general, single-step time discretization, but we primarily use first-order backwards differences for steady-state systems and second-order backwards differences for time-dependent systems.  An adaptive time step is employed, allowing for accurate time evolution with large time steps.  The nonlinear terms are treated implicitly by linearizing them using a standard, multi-iteration Newton's method similar to that given in Ref.~\onlinecite{jardinJCP08}.  For instance, the normalized heat fluxes at the $m+1$ time level, which are nonlinear functions of the macroscopic profiles, are expanded as follows:
\beq
\tld{Q}^{p+1} = \tld{Q}^{p} + \left(\mbf{y}^{p+1}-\mbf{y}^{p}\right)\pd{\tld{Q}}{\mbf{y}}\Bigg|_{\mbf{y}^{p}},
\label{eqn:qlin}
\eeq
where $p$ denotes the iteration index within each time step, $\tld{Q}\equiv(Q/pv_{th})(a/\rho)^2$, and all quantities are understood to be evaluated at time level $m+1$.  The vector $\mbf{y}$ contains the profiles of the fundamental macroscopic, time-dependent quantities in the simulation.  This consists of the two free flux functions from the Grad-Shafranov equation, $\psi$ and $I(\psi)$, as well as the species density and pressure and the species-summed toroidal angular momentum.  We are not currently evolving the magnetic equilibrium in \texttt{TRINITY}, so that $\psi$ and $I(\psi)$ are fixed in time.

Discretizing Eq.~\ref{eqn:qlin} in space, we obtain
\beq
\tld{Q}_j^{p+1} = \tld{Q}_j^{p} + \sum_k\left(\mbf{y}_k^{p+1}-\mbf{y}_k^{p}\right)\pd{\tld{Q}_j}{\mbf{y}_k}\Bigg|_{\mbf{y}_k^{p}},
\eeq
where the subscript denotes the spatial index.  In the local approximation, the fluxes depend only on the local values of macroscopic quantities and their gradients.  The above expression thus reduces to
\beq
\tld{Q}_j^{p+1} = \tld{Q}_j^{p} + \sum_k\left(\mbf{y}_k^{p+1}-\mbf{y}_k^{p}\right)\left(\pd{\tld{Q}_j}{\mbf{y}_j}+\pd{\tld{Q}_j}{\mbf{y}'_j}\frac{d\mbf{y}'_j}{d\mbf{y}_k}\right)\Bigg|_{\mbf{y}_k^{p}}.
\eeq
In \texttt{TRINITY}, we make the further simplifying assumption that the fluxes depend more strongly on profile gradients than the local values themselves.  We then neglect $\partial Q/\partial \mbf{y}$, giving us the final expression
\beq
\tld{Q}_j^{p+1} = \tld{Q}_j^{p} + \sum_k\left(\mbf{y}_k^{p+1}-\mbf{y}_k^{p}\right)\pd{\tld{Q}_j}{\mbf{y}'_j}\frac{d\mbf{y}'_j}{d\mbf{y}_k}\Bigg|_{\mbf{y}_k^{p}}.
\eeq
If this assumption is not satisfied, it affects only the rate of convergence of the solution, not its accuracy.  

In order to numerically calculate $\partial \tld{Q}_j/\partial \mbf{y}'_j$, we must employ finite differences.  This requires us to compute $\tld{Q}_j$ at multiple values of $\mbf{y}'_j$.  This is equivalent to calculating the fluxes for both the nominal profiles and for additional profiles corresponding to each gradient that must be perturbed.  The total number of flux tube calculations required during each transport time step, $N$, is given by
\beq
N = n_r \times (1+n_p),
\eeq
where $n_r$ is the number of radial grid points in the transport solver and $n_p$ is the number of macroscopic profiles being evolved.

Once the transport equations are linearized, it is straightforward to implicitly evolve them.  The expense of the implicit evolution is negligible when compared with the cost of the nonlinear turbulence calculations so that there is no need to use an approximate Jacobian.  Because inversion of the Jacobian is essentially free, a dense matrix, arising from high order spatial derivatives, is tractable.

\subsection{Schematic}
\label{sec:schematic}

To begin the numerical calculation, the initial state of the plasma must be specified.  In particular, enough information must be given to calculate the fluxes and heating from Eqs.~\ref{eqn:pflx}-\ref{eqn:heat}.   This requires local information about the magnetic equilibrium, as well as values for the macroscopic density, flow, and temperature and their gradients at each of the flux surfaces comprising the radial grid for the transport solver.  \texttt{TRINITY} is currently capable of both an analytic and numerical specification of these quantities, with experimental values taken from the publicly accessible ITER profile database.~\cite{roachNF08}  Once these quantities are obtained at each of the radial grid points, \texttt{TRINITY} calls a solver for the fluxes.  For the ion neoclassical heat flux, \texttt{TRINITY} currently uses the simplified analytic model given in Ref.~\onlinecite{changPoF82}.  All other neoclassical and classical fluxes, which are typically small compared to turbulent fluxes,~\cite{connorPPCF93,doyleNF07} are neglected.  For a more accurate treatment of neoclassical effects, one could interface to a code such as that given in Ref.~\onlinecite{belliPPCF08}, which solves Eq.~\ref{eqn:neo} directly.  This will be the subject of future work.

For the turbulent fluxes, there are interfaces within \texttt{TRINITY} to two widely-used, nonlinear gyrokinetic codes, \texttt{GS2}~\cite{kotschCPC95} and \texttt{GENE}.~\cite{jenkoPoP00,gene}  Additionally, there are options to use the IFS-PPPL model~\cite{kotschPoP95} and other simpler analytic models.  Because we are using the local approximation, the flux calculations at each radius are independent for a given transport time step.  Consequently, each flux tube calculation can be run in parallel, with the only communication occurring when the fluxes are gathered to advance the transport equations.  Since each flux tube calculation parallelizes with high efficiency to thousands of processors, our scheme with tens of flux tubes can easily scale to hundreds of thousands of processors.

Once the steady-state turbulent fluxes are computed, they are time- and flux tube-averaged and passed back to the transport solver.  The discretized transport equations are then solved to obtain updated densities, pressures, and the species-summed toroidal angular momentum.  Boundary conditions are required to obtain unique solutions.  At the outermost radius in the simulation, we fix the values of the thermodynamic profiles, typically to values taken from experiment.  As the other boundary condition, we take the product of $V'$ with the fluxes to vanish at the magnetic axis.  Within each transport time step, the equations are iterated until the relative error upon successive iterations is less than a user-specified tolerance.  In practice, we find that two iterations is sufficient to obtain accurate results.

Currently, we only consider static magnetic equilibria in \texttt{TRINITY}.  This eliminates several terms in Eqs.~\ref{eqn:n0}-\ref{eqn:p0}, and foregoes the necessity of solving Eqs.~\ref{eqn:qpsi} and~\ref{eqn:gradshaf}.  This approximation is strictly valid in the limit of $\beta=8\pi p/B^2 \gg \sqrt{m_e/m_i}$, where the magnetic geometry evolves on a resistive time which is much longer than the energy confinement time.  However, if one were to evolve the magnetic equilibrium, then the next step would be to use the parallel current obtained from Eq.~\ref{eqn:neo} in Faraday's Law (Eq.~\ref{eqn:qpsi}) to calculate $I(\psi)$.  When combined with the updated pressure gradient, this allows us to solve for $\psi$ in Eq.~\ref{eqn:gradshaf}.  With the updated thermodynamic profiles and magnetic equilibrium, we complete the feedback loop by solving for updated fluxes.  This process is repeated as many times as is necessary to evolve the macroscopic profiles beyond the time of interest.  For steady-state simulations, approximately ten to fifteen transport time steps are typically required in \texttt{TRINITY} for convergence.

\section{Simulation results}
\label{sec:results}

\begin{table}
\caption{Experimental parameters}
\begin{center}
\begin{tabular} {c c c c c}
shot & \ time (s) & \ $B_{\phi} \ (T)$ & \ $a \ (m)$ & \ $I_p$ (MA) \\
\hline \hline
JET 19649 & \ 8.7 & \ 3.12 & \ 1.16 & \ 3.05 \\
JET 42982 & \ 14.8 & \ 4.0 & \ 0.95 & \ 3.76 \\
AUG 19649 & \ 1.35 & \  2.48 & \ 0.48 & 1.0 \\
\hline
\end{tabular}
\end{center}
\label{table:shots}
\end{table}

We illustrate the utility of our multiscale model in this section by presenting numerical results from \texttt{TRINITY} simulations and comparing these results with experimental data.  We consider an L-mode discharge from JET (shot \#19649) and H-mode discharges from JET (shot \#42982) and ASDEX Upgrade (shot \#13151).  For simplicity, we consider time slices taken from approximately steady-state periods of each discharge.  Initial thermodynamic profiles and external sources used in \texttt{TRINITY} are taken during these steady-state time slices from the \texttt{TRANSP}~\cite{budnyNF92} or \texttt{ASTRA}~\cite{pereverzevIPP91} reconstructions provided in the ITER profile database.  Some key experimental parameters for these shots are given in Table~\ref{table:shots}.  In all gyrokinetic simulations, we consider electrostatic turbulence with gyrokinetic ions and electrons.

\begin{figure} 
\begin{center}
\includegraphics[height=7.5cm,angle=0]{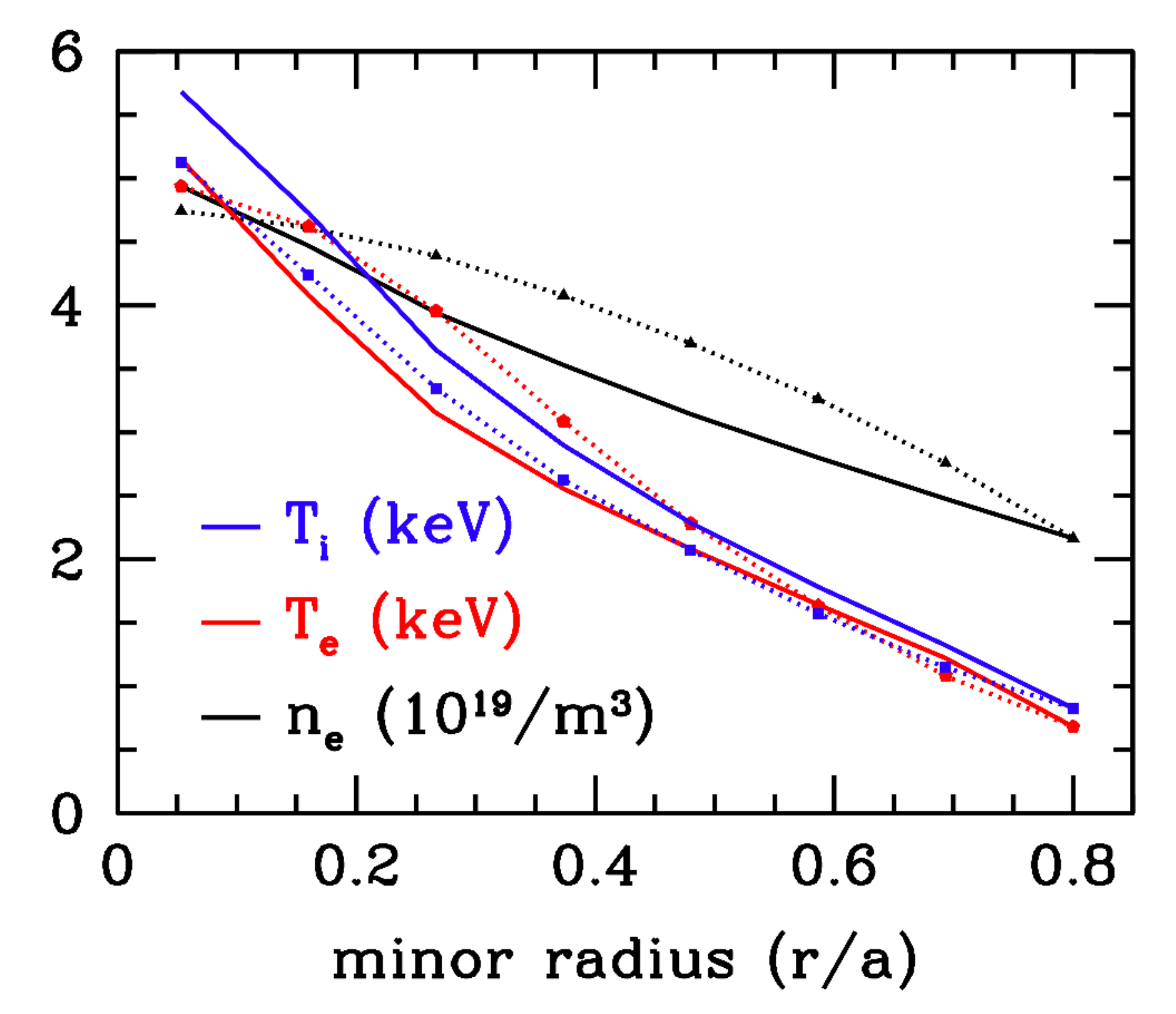}
\end{center}
\caption{Comparison of steady-state density and temperature profiles constructed from JET shot \#19649 by \texttt{TRANSP} (points and dotted lines) with those calculated in \texttt{TRINITY} (solid lines).}
\label{fig:jet19649nt}
\end{figure}

First, we consider JET shot \#19649.~\cite{baletPPCF92}  This was a standard JET L-mode pulse with $9.2 \ MW$ of neutral beam heating.  For the \texttt{TRINITY} simulation, nonlinear fluxes from \texttt{GS2} were used in the transport solver.  We used a Miller local equilibrium model~\cite{millerPoP98} for the magnetic geometry in these \texttt{GS2} simulations, with the necessary parameters taken from the ITER profile database.  In physical space, we used 16 grid points along the equilibrium magnetic field and a $40\times25$ grid in the perpendicular plane, with perpendicular box widths at the outboard midplane of 64 $\rho_i$.  In the dealiased Fourier space, this corresponds to covering $| k_{\theta} \rho_i | = 0, 0.1, 0.2, ... 0.8$ and $| k_r \rho_i | =   0, 0.1, 0.2, ..., 1.3$.  The velocity space grid consisted of $10$ velocities and $32$ pitch angles, giving approximately $10^7$ mesh points for each two-species \texttt{GS2} simulation.  These simulations employed a hyperdiffusion operator whose magnitude is scaled by the shearing rate of the turbulence, which provides a sub-grid model of the cascade of fluctuations to smaller scales.~\cite{smithPoP97, belli}  Previous tests of these kinds of sub-grid models have found that they can reduce the needed resolution, but more detailed convergence studies in the future could be useful.  Note also that because of the nonlinearities and stiffness of the transport, changes in the turbulence level of a few tens of percent at fixed temperature gradient have little effect on the final self-consistent temperature profile.

\begin{figure} 
\begin{center}
\includegraphics[height=7.5cm,angle=0]{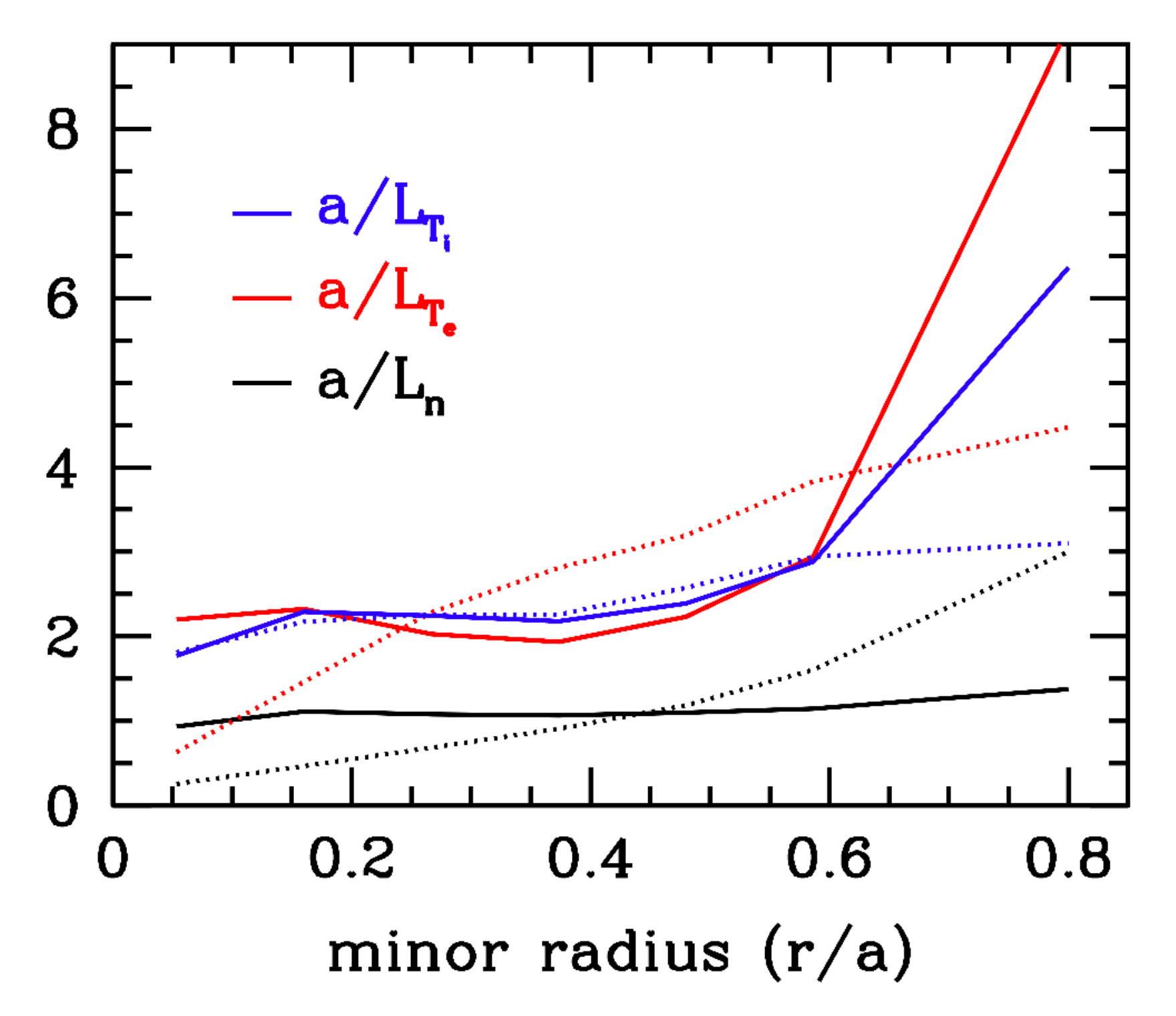}
\end{center}
\caption{Comparison of experimental (dotted lines) and simulated (solid lines) gradient scale lengths for JET shot \#19649.}
\label{fig:jet19649ntprim}
\end{figure}

Each flux tube simulation was run for $10^4$ time steps, corresponding to average physical simulation times at each radius of approximately 400-1400 $L_{T_i}/v_{th,i}$.  Electron density and ion and electron pressures were evolved at 8 radial locations, giving a total of 32 flux tube simulations per transport time step.  The total number of mesh points required for each transport time step was thus approximately $3\times10^8$.  With a total of 15 transport time steps taken, the simulation lasted approximately 4 hours on 5760 CRAY XT4 processors.  

\begin{figure} 
\begin{center}
\includegraphics[height=7.5cm,angle=0]{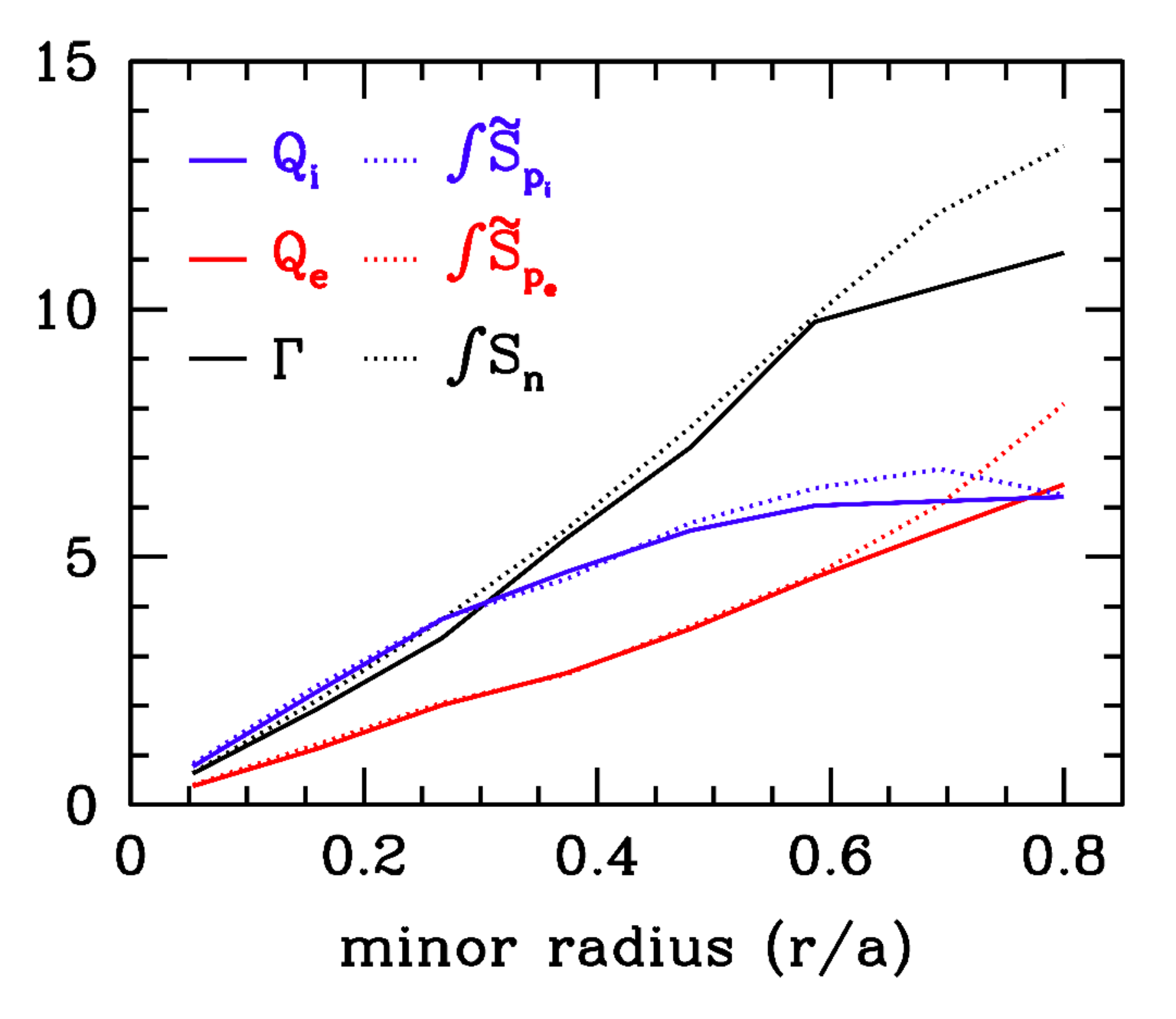}
\end{center}
\caption{Power balance for JET shot \#19649.  Solid lines are the steady-state, flux surface integrated fluxes calculated in \texttt{GS2} at the end of the \texttt{TRINITY} simulation.  Dotted lines are the volume integrals of the source terms on the right-hand side of Eqs~\ref{eqn:n0} and~\ref{eqn:p0}.  In steady-state, the solid and dotted lines should match.  The small discrepancy near the outer edge of the simulation domain is likely due to numerical inaccuracy in flux calculations at the boundary.}
\label{fig:jet19649flx}
\end{figure}

\begin{figure} 
\begin{center}
\includegraphics[height=6.9cm,angle=0]{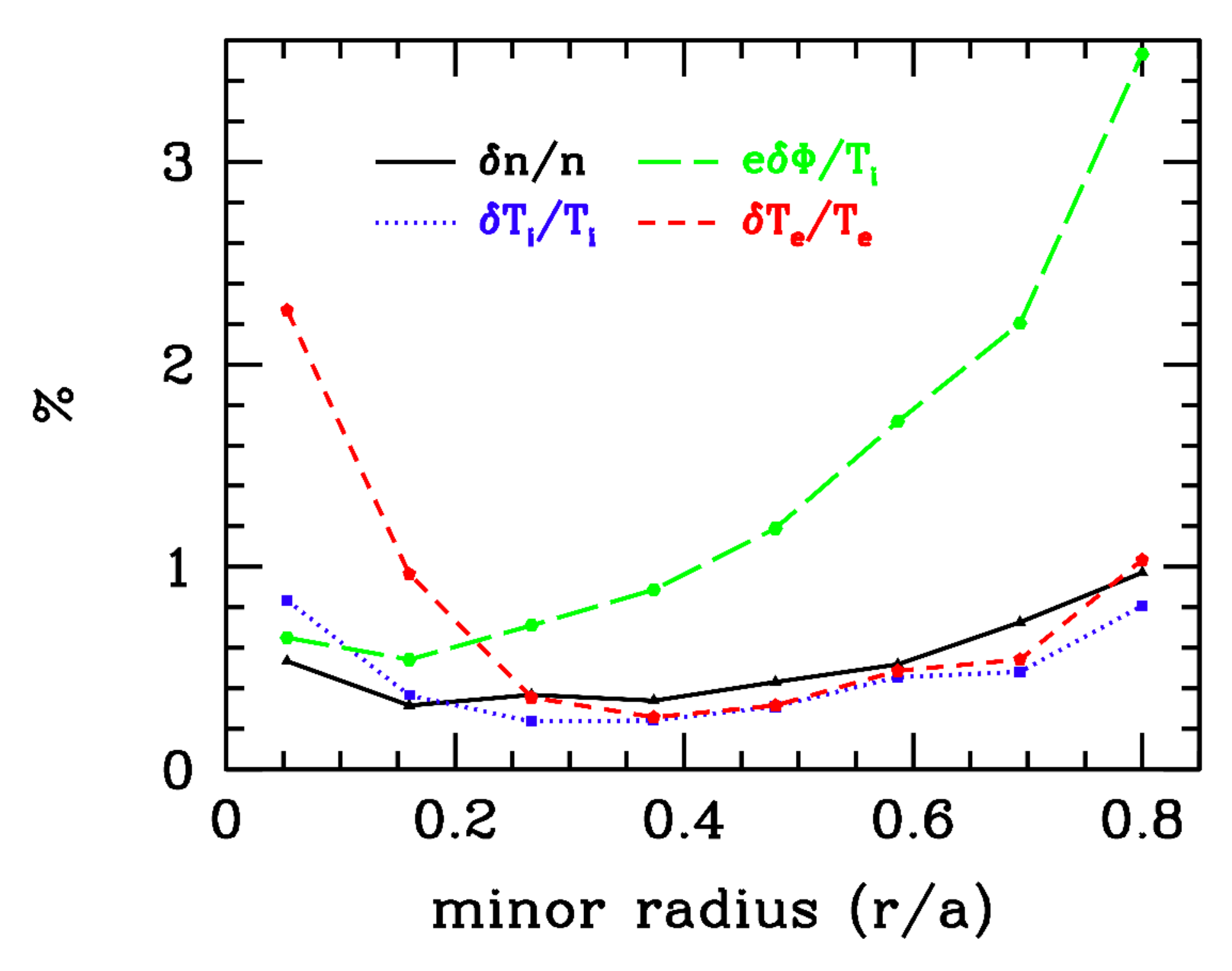}
\end{center}
\caption{Radial profiles for fluctuations of density, temperature, and electrostatic potential calculated in \texttt{GS2} for JET shot \#19649.  These fluctuations are obtained by time-averaging the instantaneous fluctuations computed in \texttt{GS2} over the steady-state period at the end of the \texttt{TRINITY} simulation.}
\label{fig:jet19649flucs}
\end{figure}

A comparison of the steady-state profiles calculated in \texttt{TRINITY} with the \texttt{TRANSP}-reconstructed experimental profiles is given in Fig.~\ref{fig:jet19649nt}.  We find good agreement for all profiles across the simulation domain ($r/a=0.053 - 0.8$), with an RMS relative error amongst all profiles ($\sqrt{(\sigma_n^2+\sigma_{T_i}^2+\sigma_{T_e}^2)/3}$) of $12\%$ (Table~\ref{table:errors}).  The total and incremental stored energies differ from \texttt{TRANSP} values by $9\%$ and $12\%$, respectively.  A comparison of the gradient scale lengths is given in Fig.~\ref{fig:jet19649ntprim}.  For $r/a<0.6$, the ion temperature gradient scale lengths from \texttt{TRINITY} and \texttt{TRANSP} match almost perfectly.  However, the electron density and temperature gradient scale lengths do not agree as well, despite reasonable profile agreement.  This illustrates a difficulty in using experimental profile measurements in standalone gyrokinetic turbulence calculations: small changes in experimental profiles can lead to significant changes in experimental gradient scale lengths, which drastically affect the turbulent fluxes.  An example of this is seen in Fig.~\ref{fig:jet19649flx}.  Here, we compare the volume integrated source terms to the flux surface integrated fluxes, which should be equal in steady state.  We see that this balance is satisfied for the self-consistent profiles obtained in \texttt{TRINITY}, but not with the profiles taken from \texttt{TRANSP}.  As an example of the sort of diagnostic information available in our multiscale gyrokinetic calculations, we show the steady-state fluctuation amplitudes for density, temperature, and electrostatic potential in Fig.~\ref{fig:jet19649flucs}.  All fluctuation amplitudes are small compared to equilibrium quantities, giving us a check on our $\delta f \ll f$ assumption.  Interestingly, we see peaked electron temperature fluctuations at the magnetic axis, while the electrostatic potential increases with radius.

\begin{table}
\caption{Analysis of \texttt{TRINITY} profile fits. $\delta_W$ and $\delta_{W_I}$ are the relative errors in total and stored energy, respectively. $\sigma$ is the RMS relative error associated with the subscripted quantity.}
\begin{center}
\begin{tabular} {c c c c c c}
shot & \ $\delta_W$ & \ $\delta_{W_I}$ & \ $\sigma_n$ & \ $\sigma_{T_i}$ & \ $\sigma_{T_e}$ \\
\hline \hline
JET 19649 & \ 0.09 & \ 0.12 & \ 0.11 & \ 0.12 & \ 0.13 \\
JET 42982 & \ 0.06 & \ 0.14 & \ 0.13 & \ 0.11 & \ 0.12 \\
AUG 19649 & \ 0.16 & \  0.29 & \ 0.13 & 0.16 & \ 0.06 \\
\hline
\end{tabular}
\end{center}
\label{table:errors}
\end{table}

Next, we consider JET shot \#42982,~\cite{hortonNF99,budnyPoP00} which achieved a record fusion energy yield of $22 \ MJ$ with 21.5 $MW$ of neutral beam heating.  The plasma in this shot was a 50-50 deuterium-tritium mixture, operating in ELMy H-mode.  Again, \texttt{GS2} was used to calculate the turbulent fluxes, with a Miller local equilibrium model for the magnetic geometry.  The perpendicular spatial grid used was the same as for the L-mode discharge, but the resolution in the remaining dimensions was increased to $24$ parallel grid points, $12$ velocities, and $40$ pitch angles.  We again employed a hyperdiffusion operator to prevent a cascade to sub-grid scales.  Each flux tube simulation was run $10^4$ time steps, with average simulation times ranging from 250-1000 $L_{T_i}/v_{th,i}$.  This time the electron density and ion and electron pressures were evaluated at 10 radial locations, resulting in 40 flux tube simulations per transport time step.  A total of 20 transport time steps were taken, and the full simulation ran just under ten hours on 8640 CRAY XT4 processors.  

\begin{figure} 
\begin{center}
\includegraphics[height=7.5cm,angle=0]{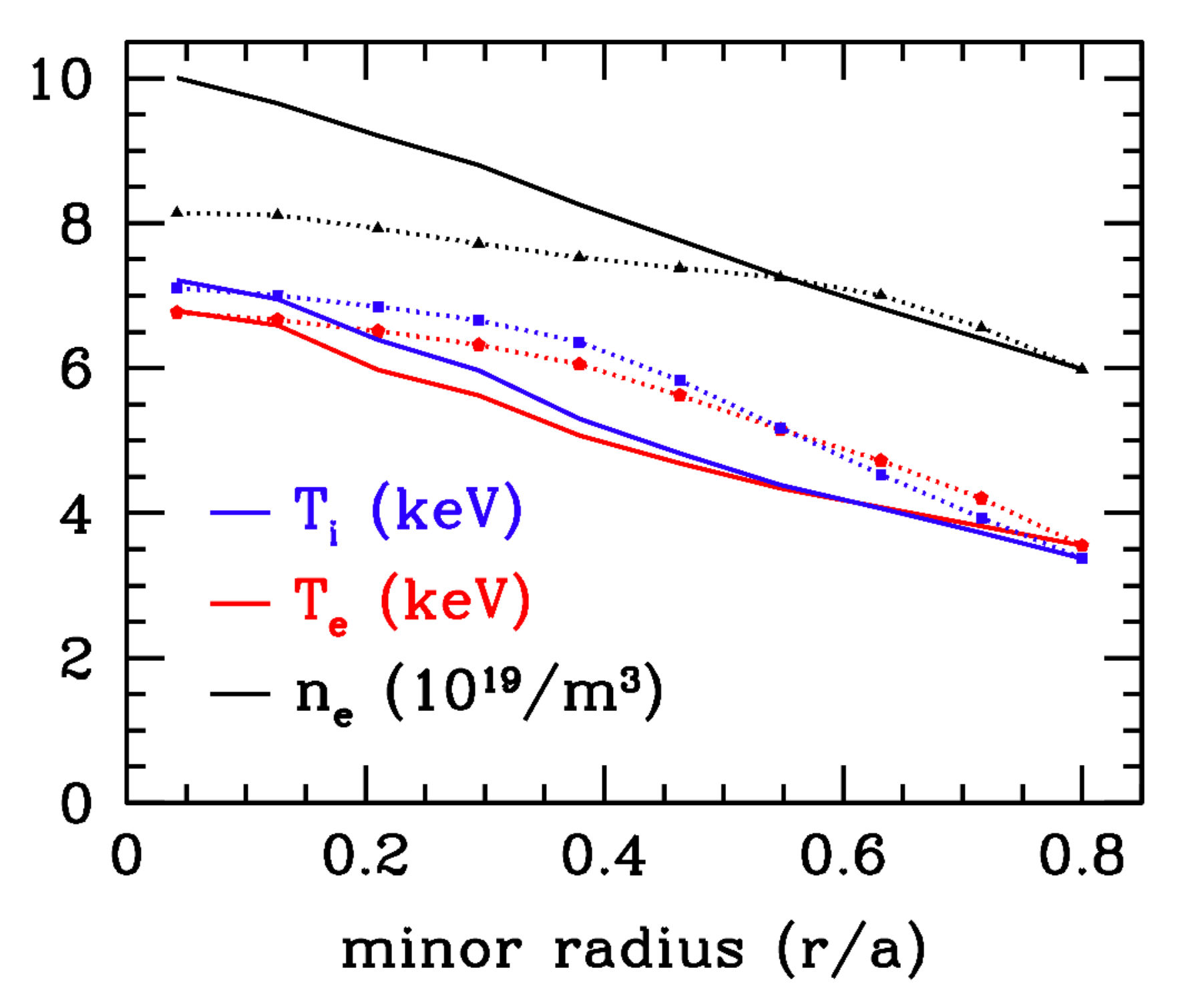}
\end{center}
\caption{Comparison of steady-state density and temperature profiles constructed from JET shot \#42982 by \texttt{TRANSP} (points and dotted lines) with those calculated in \texttt{TRINITY} (solid lines).}
\label{fig:jet42982nt}
\end{figure}

\begin{figure} 
\begin{center}
\includegraphics[height=7.5cm,angle=0]{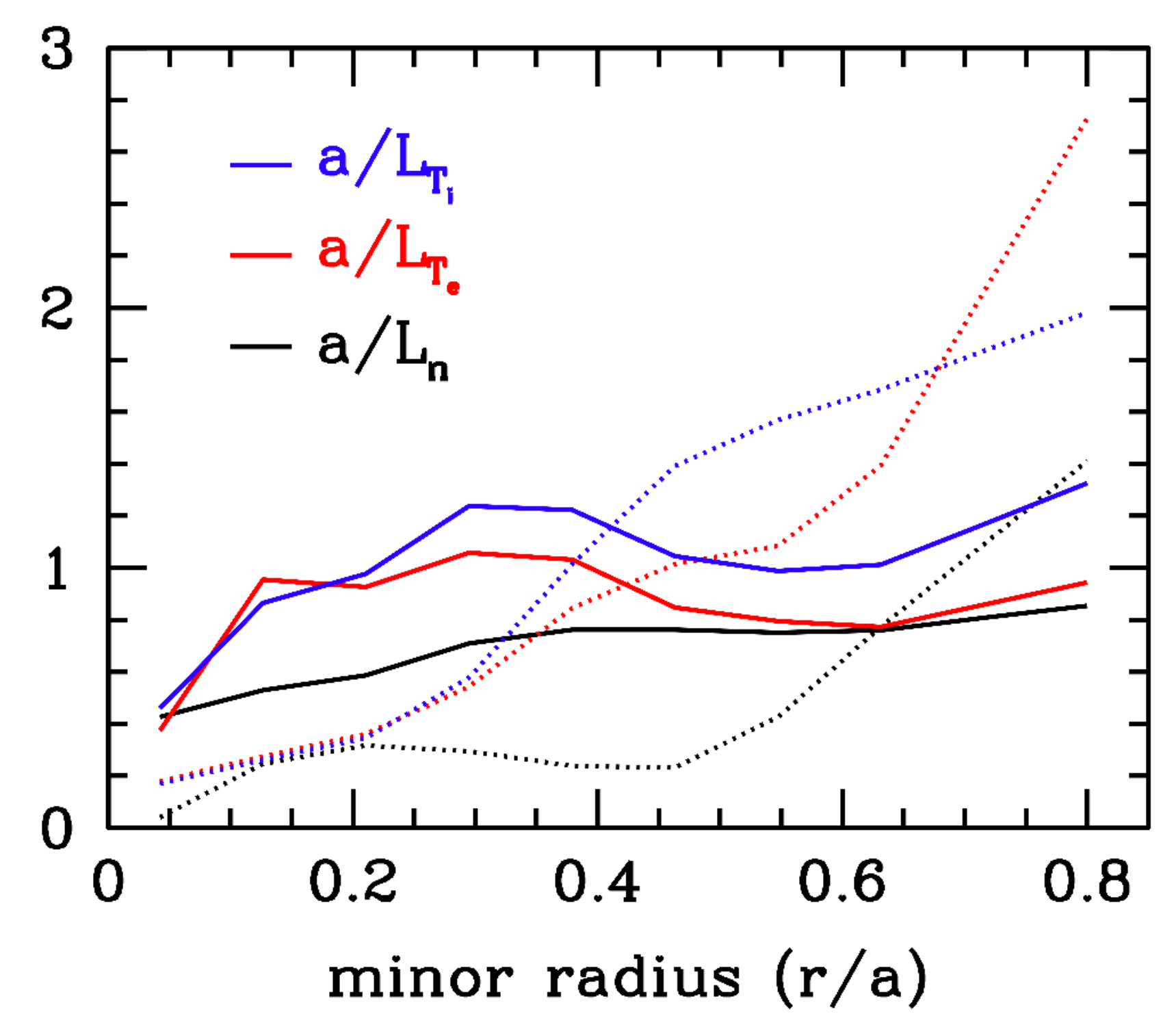}
\end{center}
\caption{Comparison of experimental (dotted lines) and simulated (solid lines) gradient scale lengths for JET shot \#42982.}
\label{fig:jet42982ntprim}
\end{figure}

\begin{figure} 
\begin{center}
\includegraphics[height=6.9cm,angle=0]{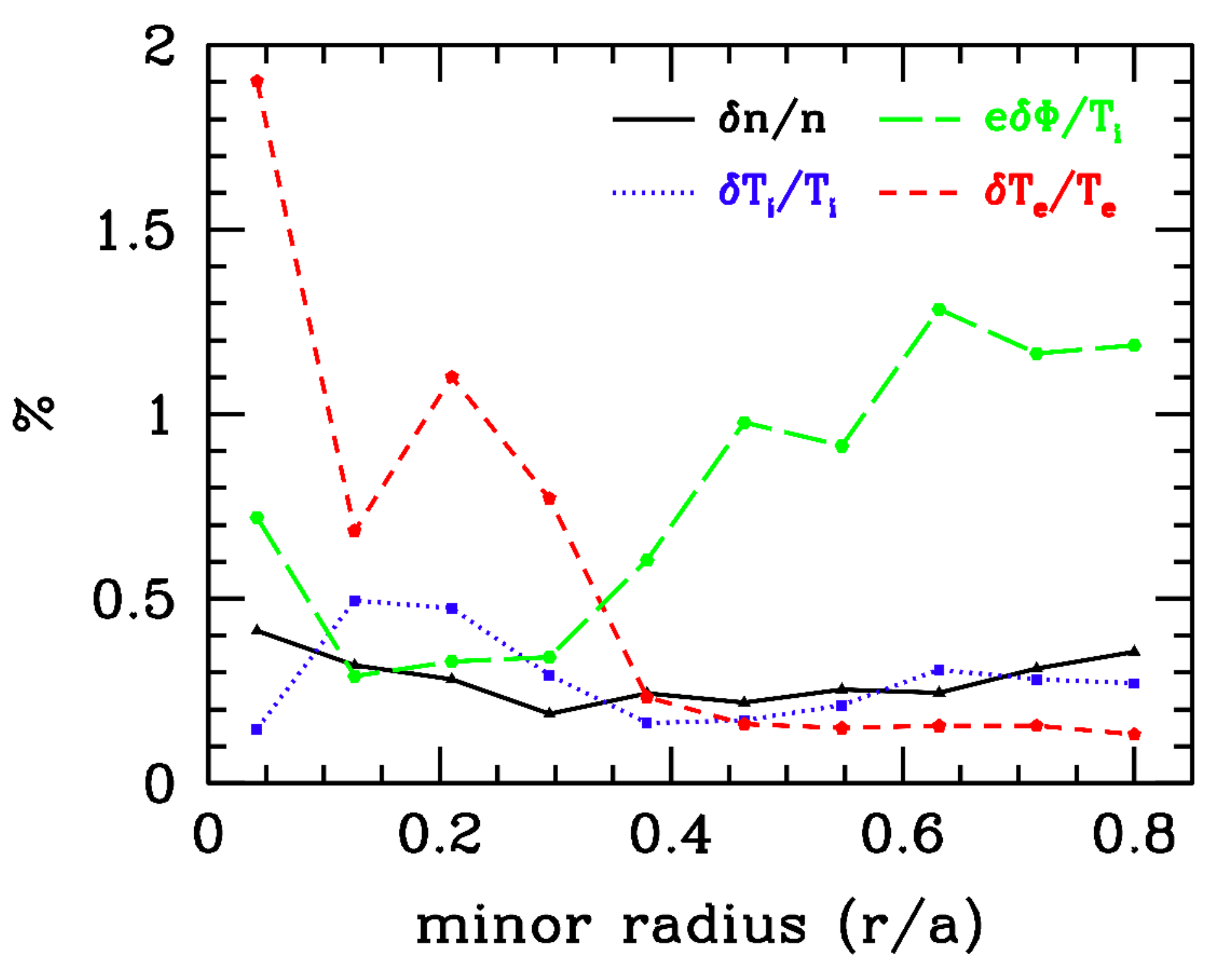}
\end{center}
\caption{Radial fluctuation profiles for density, temperature, and electrostatic potential calculated in \texttt{GS2} for JET shot \#42982.  The fluctuation levels for this H-mode discharge are generally lower than for the L-mode discharge shown in Fig.~\ref{fig:jet19649flucs}, but they exhibit the same qualitative trends in their radial profiles.}
\label{fig:jet42982flucs}
\end{figure}

As with the L-mode case, all profiles show relatively good quantitative agreement with their \texttt{TRANSP} counterparts (Fig.~\ref{fig:jet42982nt}), with an RMS relative error averaged over all profiles of $12\%$.  In this case, there appears to be a systematic over-prediction of the ion and electron heat fluxes over the outer half of the minor radius, with the profiles pinned by the critical gradient (Fig.~\ref{fig:jet42982ntprim}).  A possible explanation for this discrepancy is the fact that the radial flow shear, which has been found to have a significant effect in this discharge,~\cite{budnyEPS05} was not included in our \texttt{GS2} simulations.  This capability does exist in \texttt{GS2}~\cite{hammettDPP06,roachPPCF09} and will be included in future \texttt{TRINITY} studies with evolving toroidal angular momentum.  Again, fluctuation amplitudes are on the order of a percent of equilibrium amplitudes, with electron temperature fluctuations peaking on-axis and electrostatic potential fluctuations increasing with radius (Fig.~\ref{fig:jet42982flucs}).

\begin{figure} 
\begin{center}
\includegraphics[height=7.5cm,angle=0]{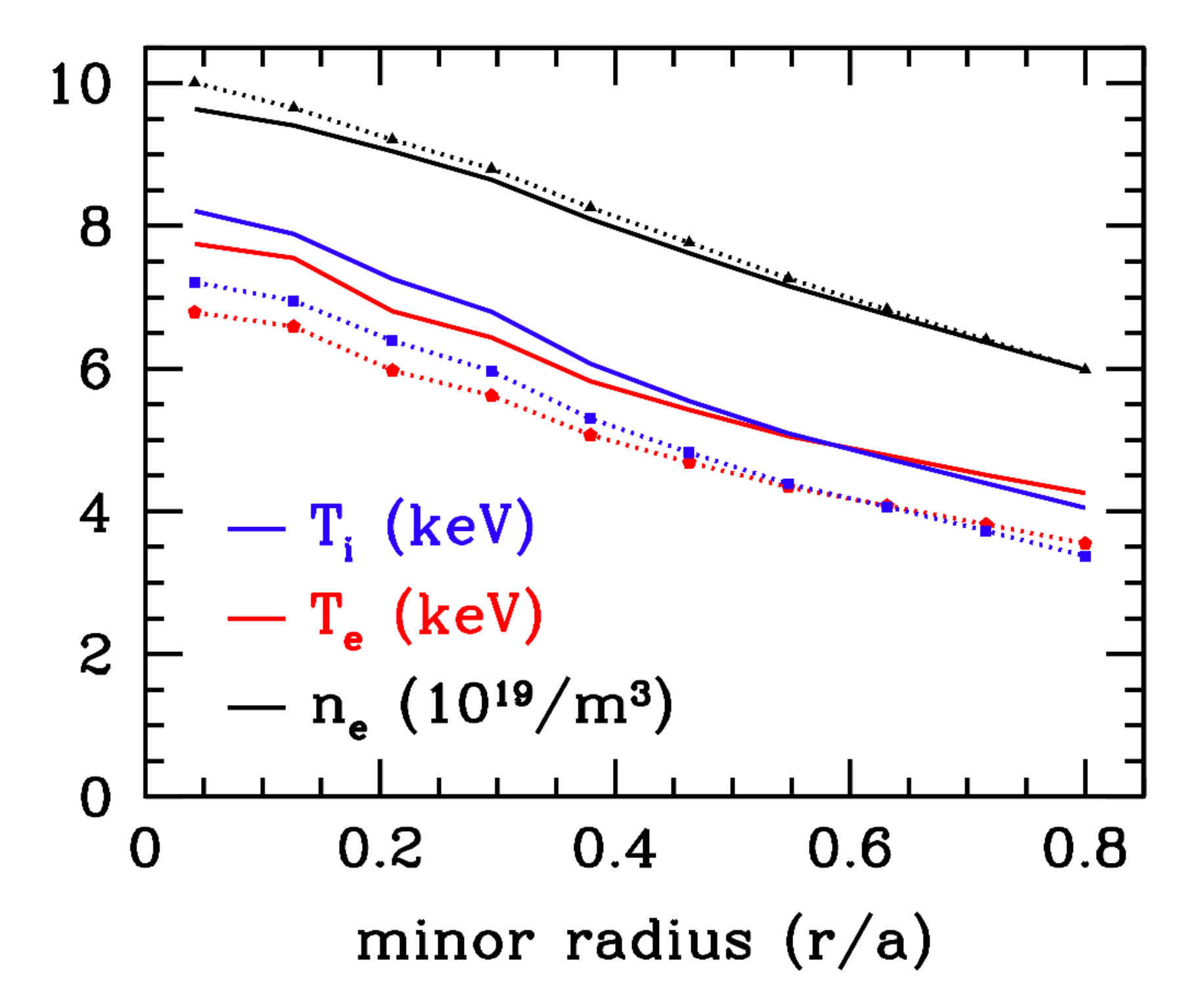}
\end{center}
\caption{Comparison of steady-state density and temperature profiles for \texttt{TRINITY} simulations of JET shot \#42982 with different edge temperatures.  The dotted lines with points correspond to a simulation using the edge temperatures reported by experiment, and the solid lines correspond to a simulation with the edge temperatures increased by $20\%$.  We see that the increased edge temperatures lead to an increase in the temperatures near the magnetic axis of about $14\%$.}
\label{fig:edge42982nt}
\end{figure}

\begin{figure} 
\begin{center}
\includegraphics[height=7.5cm,angle=0]{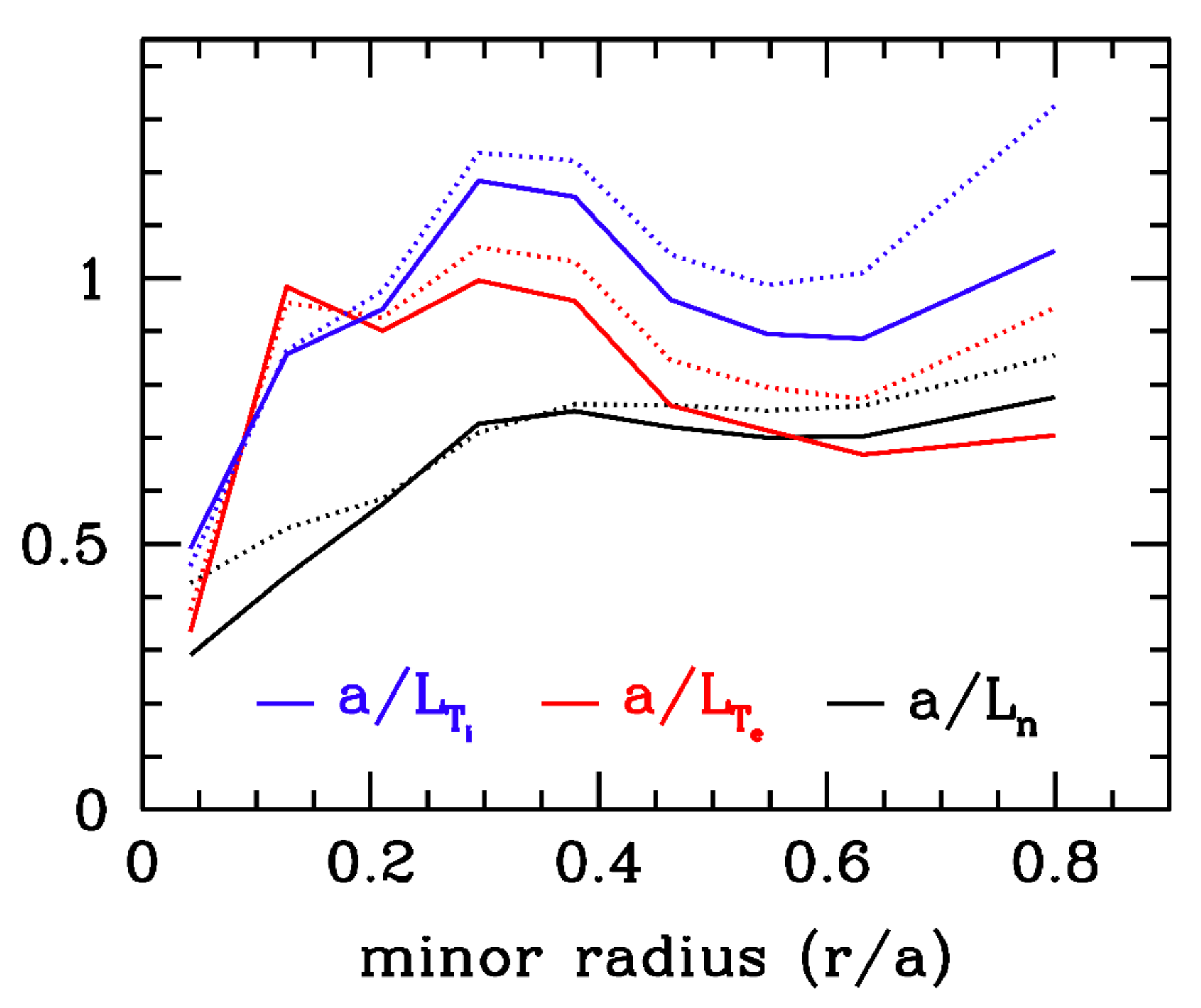}
\end{center}
\caption{Comparison of gradient scale lengths for two different \texttt{TRINITY} simulations of JET shot \#42982.  Dotted lines correspond to a simulation with edge temperatures taken from experiment, and solid lines correspond to a simulation with the edge temperatures increased by $20\%$.  While the gradient scale length profiles are quite similar, the case with higher edge temperature leads to lower gradient scale lengths near the edge, where the profiles are likely less stiff.}
\label{fig:edge42982ntprim}
\end{figure}

\begin{figure} 
\begin{center}
\includegraphics[height=6.9cm,angle=0]{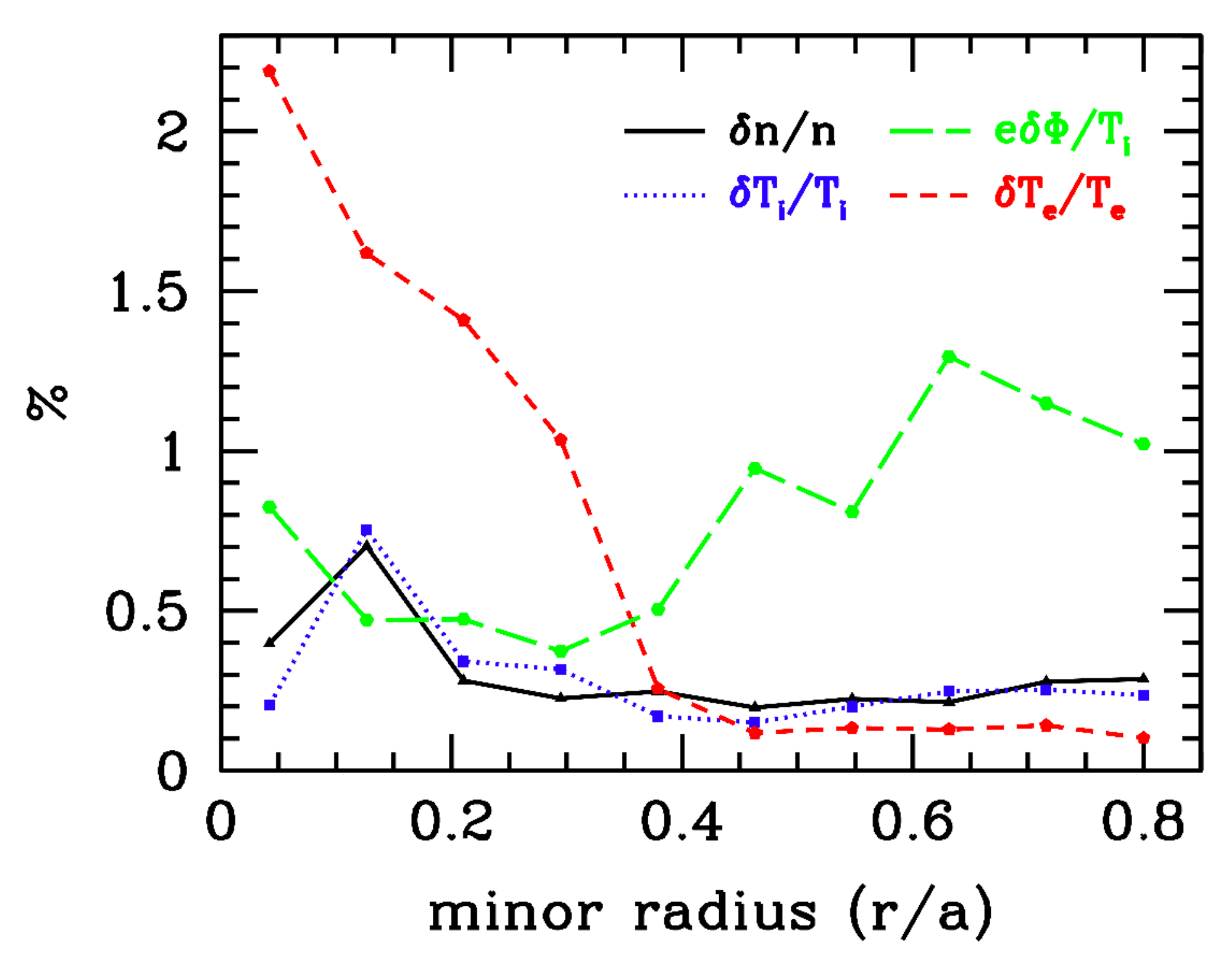}
\end{center}
\caption{Radial fluctuation profiles for density, temperature, and electrostatic potential calculated in \texttt{GS2} for JET shot \#42982 with increased edge temperatures.  They are quite similar to the case with edge temperatures taken from experiment.}
\label{fig:edge42982flucs}
\end{figure}

In order to study the effect of the edge temperature on the profiles, we repeated this simulation with electron and ion edge temperatures increased by $20\%$.  The results are shown in Figs.~\ref{fig:edge42982nt}-\ref{fig:edge42982flucs}.  We see that the $20\%$ increase in edge temperature leads to an increase of approximately $14\%$ at the magnetic axis, as expected from the stiff profiles indicated in Fig.~\ref{fig:jet42982ntprim}.  The gradient scale lengths are similar to the base case across most of the minor radius, with the only significant discrepancies occurring near the edge where the stiffness of the profiles is less pronounced (Fig.~\ref{fig:edge42982ntprim}).  Fluctuation levels also do not differ significantly from those obtained for the base case (Fig.~\ref{fig:edge42982flucs}).

Finally, we consider ASDEX Upgrade shot \#13151,~\cite{tardiniNF02} which was an ELMy H-mode discharge with $5 \ MW$ of neutral beam heating.  Here, we used \texttt{GENE} to calculate the turbulent fluxes, with a numerical magnetic equilibrium generated by \texttt{TRACER}.~\cite{xanthopoulosPoP06}  We used $16$ parallel grid points and a $64\times48$ grid in the perpendicular spatial plane, with perpendicular box widths of approximately 64 $\rho_i$ at the outboard midplane.  The velocity space was sampled with $32$ parallel velocities and $8$ magnetic moments, resulting in a total of approximately $2.5\times10^7$ mesh points for each \texttt{GENE} simulation.  A linearized Landau-Boltzmann operator was used to model the effect of collisions.~\cite{merz}  The number of time steps taken in each flux tube simulation varied from about 3-4$\times10^4$ depending on radial location, corresponding to simulation times of approximately 400-1000 $L_{T_i}/v_{th,i}$.  Electron density and ion and electron pressures were again evolved at 8 radial locations, resulting in 32 flux tube calculations per transport time step.  Within each transport time step, the total mesh points required was thus $\approx8\times10^8$.  Due to the increased spatial resolution in the perpendicular domain and the use of a physical collision operator, the \texttt{GENE} simulations took somewhat longer than the earlier \texttt{GS2} simulations.  In total, the simulation took just under 24 hours for 16 transport time steps on 16384 CRAY XT4 processors.

\begin{figure} 
\begin{center}
\includegraphics[height=7.5cm,angle=0]{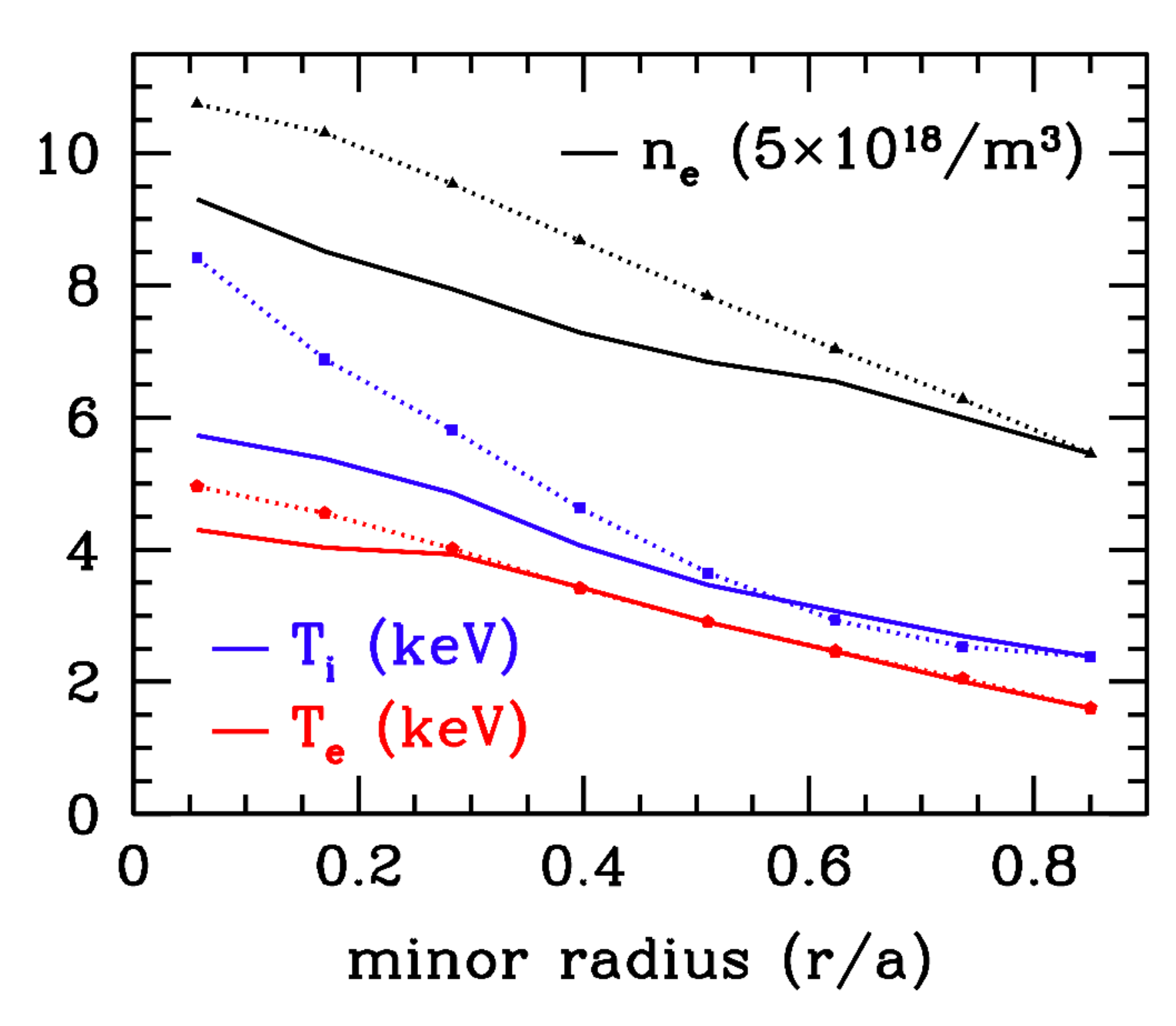}
\end{center}
\caption{Comparison of steady-state density and temperature profiles constructed from ASDEX Upgrade shot \#13151 by \texttt{ASTRA} (points and dotted lines) with those calculated in \texttt{TRINITY} (solid lines).}
\label{fig:aug13151nt}
\end{figure}

\begin{figure} 
\begin{center}
\includegraphics[height=7.5cm,angle=0]{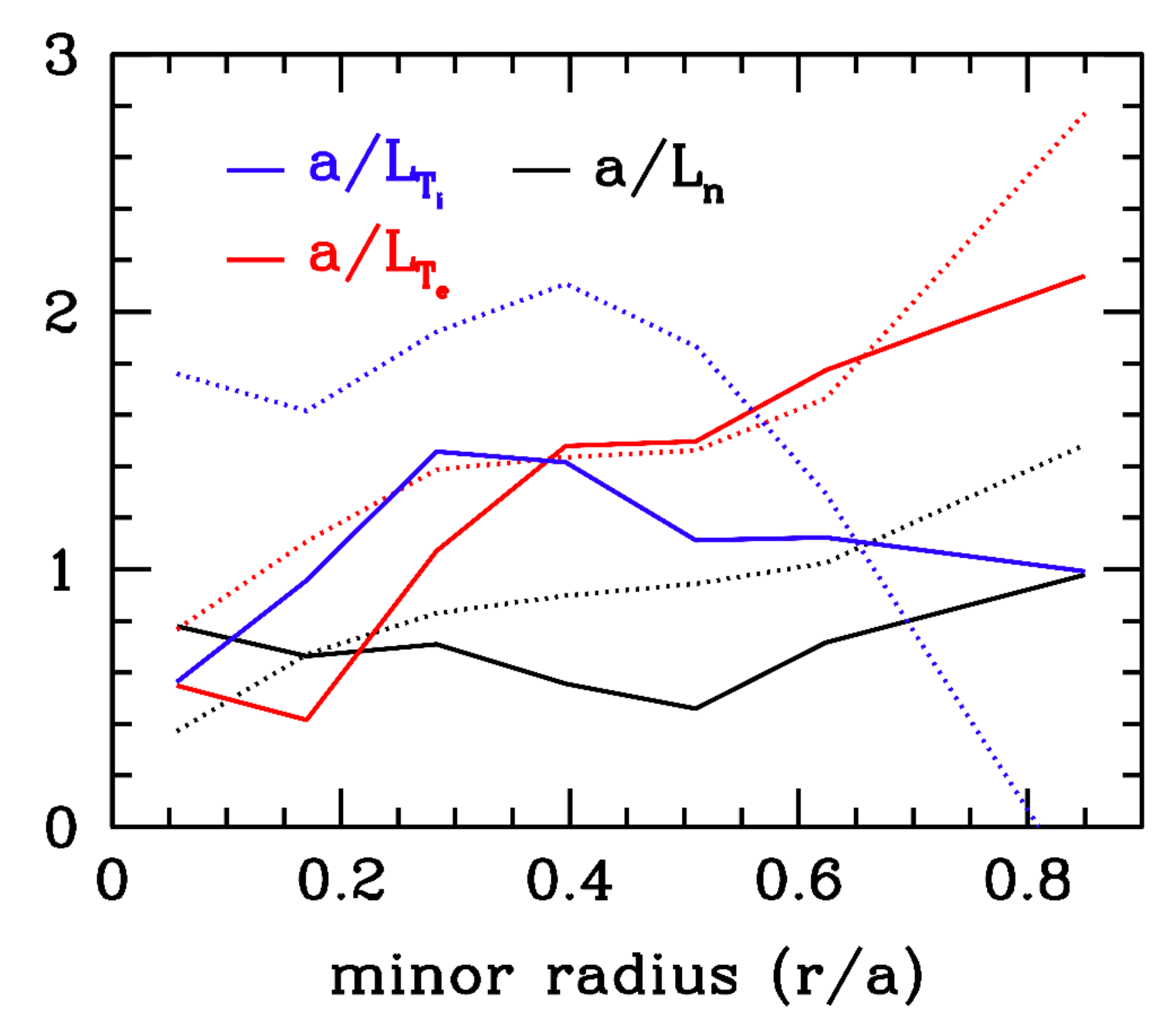}
\end{center}
\caption{Comparison of experimental (dotted lines) and simulated (solid lines) gradient scale lengths for ASDEX Upgrade shot \#13151.}
\label{fig:aug13151ntprim}
\end{figure}

The electron density and temperature agree relatively well across the minor radius, but the ion temperature is under-predicted near the magnetic axis (Fig.~\ref{fig:aug13151nt}).  This is reflected in the profile gradient scale lengths shown in Fig.~\ref{fig:aug13151ntprim}.  The RMS relative error averaged over profiles is nonetheless only $12\%$.  A lack of flow shear in the calculation of the turbulent fluxes is again a possible explanation for the discrepancy in ion temperature near-axis.  Another possibility is the fact that no MHD model was used for the $q$ profile computed by \texttt{ASTRA}, which resulted in $q<1$ inside $r/a\approx0.4$, with $q(0)<0.5$.  Consequently, there is significant uncertainty in modeling the system in this region.

\section{Conclusions}
\label{sec:conclusion}

We have presented in this paper a complete theoretical and numerical model for the interaction of micro- and macro-physics in axisymmetric fusion devices.  This model arises from a rigorous asymptotic expansion of the full Maxwell-Boltzmann system.  The ordering assumptions used in the expansion were given in \secref{sec:theory}, along with the resulting closed system of equations for the evolution of the macroscopic thermodynamic profiles and magnetic geometry (Eqs.~\ref{eqn:n0}-\ref{eqn:gradshaf}).  These profiles depend on fluxes and heating (Eqs.~\ref{eqn:pflx}-\ref{eqn:heat}) arising from classical, neoclassical, and turbulent dynamics, requiring solution of the neoclassical and gyrokinetic equations (Eqs.~\ref{eqn:neo} and~\ref{eqn:gk}).  In order to evolve the magnetic geometry and close the system, the toroidal component of Ampere's Law (Eq.~\ref{eqn:qpsi}) and the Grad-Shafranov equation (Eq.~\ref{eqn:gradshaf}) also have to be solved.

In \secref{sec:numerics}, we described the numerical scheme used in \texttt{TRINITY} to solve the multiscale gyrokinetic system of equations.  The key idea in the approach is to use scale separations in space and time to embed a fine mesh for turbulent dynamics in a coarse grid for the transport solver.  A local, nonlinear gyrokinetic code is used to calculate the turbulent fluxes, which are then passed to the transport solver.  The macroscopic thermodynamic profiles are then evolved in a manner consistent with the small $\rho_*$ limit of the orderings used to derive the nonlinear gyrokinetic equation.  Thus, \texttt{TRINITY} can be used to simulate time-dependent experimental phenomena.  A Newton method is used to devise an implicit time stepping algorithm, allowing for large time steps characteristic of the confinement time.  With updated pressure profiles, one could then couple to solvers for the toroidal component of Faraday's Law and the Grad-Shafranov equation to evolve the magnetic geometry.

Simulation results from \texttt{TRINITY} are provided in \secref{sec:results}, with comparisons to JET and ASDEX Upgrade plasmas.  Relatively good agreement is found for all evolved profiles (density and electron/ion pressures), with an RMS relative error averaged over all profiles of $12\%$.  Fluctuation levels for density, temperature, and electrostatic potential are on the order of a percent across the minor radius, in general agreement with experimental evidence.

While currently capable of faithfully simulating a range of interesting experimental conditions, there are several useful additions that could be made to the numerical model implemented in \texttt{TRINITY}.  Coupling to a neoclassical code which solves Eq.~\ref{eqn:neo} would provide more accurate neoclassical fluxes, which can be important in certain experimental regimes (for instance, when strong shear flow partially suppresses turbulent flux levels).  Additionally, it would allow for the calculation of the parallel current necessary to evolve the safety factor and the poloidal flux.  Coupling to a Grad-Shafranov solver would then allow for a study of the effects associated with evolving magnetic geometry.  Additionally, coupling to a linear MHD code could be useful in monitoring macroscopic profiles to ensure that MHD stability boundaries are not crossed.

In conclusion, we emphasize that the multiscale gyrokinetic model presented here is not meant to be a comprehensive model for all physics present in a tokamak discharge.  Instead, it is meant to be used as a tool for studying the self-consistent interaction between micro-turbulence and macroscopic profiles.  The ultimate goal of this approach is to obtain both a better qualitative and quantitative understanding of this interaction in order to enhance our ability to suppress turbulence and improve confinement in fusion experiments.

\begin{acknowledgments} 
The authors would like to thank A. A. Schekochihin, S. C. Cowley, G. G. Plunk, and F. Parra for useful discussions.  
M.B. and W.D. were supported in part by the U.S. DOE Center for Multiscale Plasma Dynamics.  M.B. was also supported by the Oxford-Culham Fusion Research Fellowship.  I.G.A.\ was supported by a CASE EPSRC studentship in association with UKAEA Fusion 
(Culham).  M.B., I.G.A., W.D., and G.W.H.\ would also like to thank the Leverhulme Trust (UK) 
International Network for Magnetized Plasma Turbulence for travel support.  Computing time for this research was provided in part by EPSRC grant EP/H002081/1.
\end{acknowledgments}


\end{document}